\documentclass[12pt]{article}
\usepackage{epsfig}
\usepackage{color}
\usepackage{braket}

\newcommand{\be}{\begin{equation}}
\newcommand{\ee}{\end{equation}}

\newcommand{\bea}{\begin{eqnarray}}
\newcommand{\eea}{\end{eqnarray}}
\newcommand{\beq}{\begin{equation}}
\newcommand{\eeq}{\end{equation}}

\def\fun#1#2{\lower3.6pt\vbox{\baselineskip0pt\lineskip.9pt
\ialign{$\mathsurround=0pt#1\hfil##\hfil$\crcr#2\crcr\sim\crcr}}}

\begin{document}

\title{
Pentaquarks and strange tetraquark mesons
}
\author{
V.V. Anisovich$^+$, M.A. Matveev$^+$,
A.V. Sarantsev$^{+ \diamondsuit}$,  A.N. Semenova$^+$
}

\date{today}
\maketitle

\begin{center}
{\it
$^+$National Research Centre ''Kurchatov Institute'':
Petersburg Nuclear Physics Institute, Gatchina, 188300, Russia}

{\it $^\diamondsuit$
Helmholtz-Institut f\"ur Strahlen- und Kernphysik,
Universit\"at Bonn, Germany}

\end{center}

\begin{abstract}
We consider the interplay of the pentaquark states and strange
tetraquark states in the decay  $\Lambda^0_b\to K^-J/\psi p$.
Possible existence of ($cs\bar c\bar u$)-states
is taken up and their mani\-festation in the $K^-J/\psi $-channel is discussed.
It is emphasised that these exotic mesons can imitate broad bumps in the $pJ/\psi $-channel.
\end{abstract}


\section{Introduction}

The LHCb collaboration \cite{LHCbB} claims two candidates for pentaquark states
which are seen in the decay $\Lambda^0_b\to K^-J/\psi\, p$.
These are narrow peak $5/2^?(4450\pm 4)$ with a width $\Gamma=(39\pm 24)$ MeV
and a broad state  $3/2^?(4380\pm 38)$ with
$\Gamma=(205\pm 94)$ MeV.
In the $(K^-\,p)$-channel a set of the observed $\Lambda$-states \cite{PDG} is taken into
account in the fit procedure of ref. \cite{LHCbB} while an existence of resonances in the
$(K^-J/\psi)$-spectrum is turned down. But
in the quark-diquark schemes the adoption of the pentaquark states means an
adoption of the tetraquark states as well. Then the $(K^-J/\psi)$-spectrum should be
filled up by exotic mesons, supposing these exotic states exist.

In this note we emphasize that the strange exotic mesons with hidden charm,
$(c\bar c\, s\bar u)$, (masses and spins have been estimated in \cite{ala-1}) can decay
by process $(c\bar c\, s\bar u)\to K^-J/\psi$ thus imitating contributions from the
$pJ/\psi$-channel.

\begin{figure}
\centerline{\epsfig{file=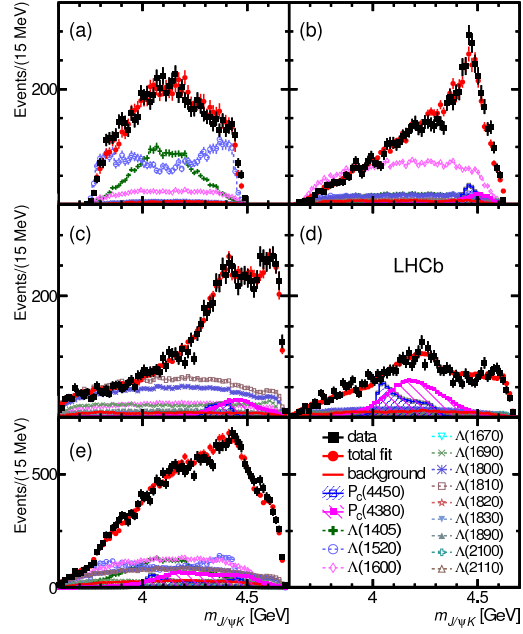,height=0.50\textwidth}}
\caption{\label{f1}
Fig.11 of the ref. [1]: Projections onto $m_{J/\psi K}$ in various intervals of
$m_{Kp}$ for the reduced model fit (cFit) with two $P^+_c$ states of $J^P$ equal to
$3/2^-$ and $5/2^+$: (a) $m_{Kp}< 1.55$ GeV, (b) $1.55<m_{Kp}< 1.70$ GeV,
(c) $1.70<m_{Kp}< 2.00$ GeV, (d) $2.00<m_{Kp}< 2.00$ GeV. The data are shown as (black)
squares with error bars, while the (red)
circles show the result of the
fit. The individual resonances are given in the legend
. }
\end{figure}

\section{Exotic meson states}

Strange exotic mesons with hidden charm were discussed in \cite{ala-1} as composite diquark-antidiquark systems.
In terms of the quark model we have found the masses and spins of such states.
Results for the $(cs\cdot\bar c\bar s)$ and $(cq\cdot\bar c\bar s)$ systems read:
\be \label{33}
\begin{tabular}{l|l|l||l|l|l}
&$(cs)\cdot(\bar c\bar s)$ & mass  & & $(cq)\cdot(\bar c\bar s)$& mass  \\
$J^{PC}$&$I=0$&MeV  & $J^{P}$& $ I=1/2$& MeV   \\
\hline
$2^{++}$& $ A_{(cs)}\cdot A_{(\bar c\bar s)}$&$4630\pm 50$ &
$2^{+}$& $ A_{(cq)}\cdot A_{(\bar c\bar s)}$& $4530\pm 50$\\
$1^{+-}$& $ A_{(cs)}\cdot A_{(\bar c\bar s)}$&$4400\pm 50$ &
$1^{+}$& $ A_{(cq)}\cdot A_{(\bar c\bar s)}$& $4300\pm 50$\\
$0^{++}$& $ A_{(cs)}\cdot A_{(\bar c\bar s)}$&$4280\pm 50$ &
$0^{+}$& $ A_{(cq)}\cdot A_{(\bar c\bar s)}$& $4180\pm 50$\\
$1^{++}$& $ [A_{(cs)}\cdot S_{(\bar c\bar s)}]_{sym}$&$4380\pm 50$ &
$1^{+}$& $ A_{(cq)}\cdot S_{(\bar c\bar s)}$& $4280\pm 50$\\
$1^{+-}$& $ \{S_{(cs)}\cdot A_{(\bar c\bar s)}\}_{asym}$&$4330\pm 50$ &
$1^{+}$& $ S_{(cq)}\cdot A_{(\bar c\bar s)}$& $4230\pm 50$\\
$0^{++}$& $ S_{(cs)}\cdot S_{(\bar c\bar s)}$&$4140\pm 50$ &
$0^{+}$& $ S_{(cq)}\cdot S_{(\bar c\bar s)}$& $4040\pm 50$\\
 \end{tabular}
\ee
Here $A_{(cs)}$ and $ S_{(\bar c\bar s)}$ are axial ($1^+$) diquark and scalar ($0^+$)
antidiquark correspondingly. The masses of the states $(A_{(cs)}\cdot A_{(\bar c\bar s)})$,
$(A_{(cs)}\cdot S_{(\bar c\bar s)})$, $(S_{(cs)}\cdot S_{(\bar c\bar s)})$ are
determined basing on the Belle \cite{belle-ss}, CDF \cite{cdf-ss}, CMS \cite{CMS-ss},
D0 \cite{D0-ss} data by the diquark - antidiquark model \cite{ala-1}. This model
treats the exotic meson states
 as two-component composite systems with\\
(i) diquark-antidiquark component $(cs)\cdot(\bar c\bar s)$, \\
(ii) meson-meson component $(c\bar s)\cdot(s\bar c)$. \\
The masses of exotic mesons with open strangeness,
$(cq)\cdot(\bar c\bar s)$ and  $(cs)\cdot(\bar c\bar q)$, are fixed by
strange quark weighting, it is accepted to be
$\Delta m_s =100$ MeV (this point is discussed in \cite{ala-1}).

\section{Interplay of of exotic meson and baryon states}

The $(K^-J/\psi)$-spectra  measured in \cite{LHCbB}
 are shown on Fig. \ref{f1} (figure 11 of ref. \cite{LHCbB}),
the mass distributions have structures which nicely coincide with masses of the
$(cq)\cdot(\bar c\bar s)$-states shown in the last column of eq. (\ref{33}).
Let us emphasize that Fig. \ref{f1}d demonstrates data events from the mass region
with a minimal contribution of the $\Lambda$-resonances thus providing argument
for search
of resonances in the $(K^-J/\psi)$ channel.

We observe a crossing of contributions of the baryon state  $3/2^?(4380\pm 38)$ and
meson states with $J^P=0^+,1^+$ in the mass interval
$M_{(cq)\cdot(\bar c\bar s)}\sim (4180-4300)$ MeV,
and that require a study of their interplay.
The meson states can imitate a signal of the broad baryon resonance at mass near 4380 MeV.

\section{Conclusion}

A wide discussion accompanies the observation of structures in the
$p\,J/\psi $ channel
\cite{ala-2,maia-polo,mikh,liu-wang,meis-olle,miro,he,lebe,srr,x-meis,kuba-wolo,wang,he2,burns}.
In line with
\cite{ala-2} we discuss a version which suggests that
the narrow peak
 $5/2^?(4450\pm 4)$ is the genuine pentaquark state with $J^P=5/2^-$
 and the broad bump  $3/2^?(4380\pm 38)$
is mainly resulted due to exotic mesons $1^+$ in the $(K^-J/\psi)$ channel
with masses $\sim 4189-4300$ MeV.
We conclude to study the broad pentaquark states one needs to determine
the character of the irregularities in the $(K^-J/\psi)$ channel.

We are grateful to Y.I. Azimov and J. Nyiri for helpful discussions.
 The work was supported by grants RSGSS-4801.2012.2,
 RFBR-13-02-00425 and RSCF-14-22-00281 .

\end{document}